\documentclass[final]{svjour2}
\usepackage{graphicx}
\usepackage{rotating}
\usepackage{amssymb}
\usepackage{amsmath}
\usepackage{mathptmx}
\usepackage{bm}
\usepackage[numbers]{natbib}
\newcommand{\abbrev}[1]{\textsc{#1}}
\makeatletter
\journalname{}

\bibpunct{}{}{,}{s}{}{,}

\begin{document}

\title{A microscopic description of vacancies in solid $^{\bf 4}$He}

\author{  R. Rota$^1$ \and  Y. Lutsyshyn$^2$ \and   C. Cazorla$^3$ \and  
  J. Boronat$^1$ }

\institute{1:Departament de F\'\i sica i Enginyeria Nuclear, Campus Nord
B4-B5,
\\ Universitat Polit\`ecnica de Catalunya, 08034 Barcelona, Spain\\
\email{jordi.boronat@upc.edu}
\\ 2: Institut f\"{u}r Physik, Universit\"{a}t Rostock, 18051 Rostock, Germany
\\ 3: Institut de Ci$\grave{e}$ncia de Materials de Barcelona (ICMAB-CSIC), 
08193 Bellaterra, Spain }

\date{\today}

\maketitle


\begin{abstract}

The changes that vacancies produce in the properties of hcp solid $^4$He are
studied by means of quantum Monte Carlo methods. Our results show that the
introduction of vacancies produces significant changes in the behavior of solid
$^4$He, even when the vacancy concentration is very small. We show that there is 
an onset
temperature where the properties of incommensurate $^4$He change significantly.
Below this temperature, we observe the emergence of off-diagonal long range order
and a complete spatial delocalization of the vacancies. This temperature
is quite close to the temperature where non-classical rotational inertia has been 
experimentally observed. Finally, we report results on the influence of vacancies
in the elastic properties of hcp $^4$He at zero temperature.

\end{abstract}

\section{Introduction}
The counterintuitive concept of supersolidity can only be understood in
terms of extremely quantum matter. The simultaneous existence of spatial solid
order, characteristic of the solid state, and superfluidity, property that
in principle requires of mass movement without friction, is hardly
understandable in usual solids. The natural candidate for this fascinating
possibility has been along the time always the same, solid $^4$He. The solid phase
of $^4$He is not a normal solid, in the classical meaning of this term.
$^4$He atoms have so small mass and shallow interaction that 
its stable condensed phase, in the limit of zero temperature, is
a liquid and thus a finite pressure is required to crystallize it. Even in
the crystal phase, the zero-point motion of the $^4$He atoms is not fully
depressed as it is explicitly quantified by its large Lindemann ratio and
excidingly large kinetic energy, with respect to typical classical values.
Furthermore, the exchange frequency, which is absolutely absent in
classical solids, is low but not zero. Altogether makes the study of a
quantum solid a very interesting topic, the possibility of a supersolid
scenario being one of its more stimulating possibilities.

The recent experimental findings by Kim and Chan~\cite{kim04a,kim04b} 
on the existence of a
non-zero fraction of non-classical moment of inertia (NCRI) in torsional
oscillator measurements have revived this topic that emerged in the past as
an hypothetical theoretical conjecture. This unexpected result has been
corroborated by other laboratories, although the dispersion in the size of the
effect is large~\cite{balibar}. From the very beginning, it was clear that the way 
in which the crystal
is produced, the annealing during its growth and the purity of the $^4$He
sample were, among others, relevant parameters which introduce changes of
the NCRI fraction of an order of magnitude. The decoupling of a part of the
mass, measured in the torsional oscillator, is observed at an onset
temperature which also depends on the particular conditions of the
experiment, but its fluctuation is sizeably smaller than the one obtained
for the NCRI fraction (superfluid fraction). Very close to the onset
temperature for supersolidity it has been observed an increase in the shear
modulus of hcp $^4$He, its temperature dependence being similar in shape to the one 
of NCRI~\cite{day07}. The interplay between this anomalous elastic behavior and the NCRI
effect observed in torsional oscillators has been object of theoretical
debate: it seems clear that part of the NCRI effect can be attributed to
elasticity but probably not all.

To date, there is not a complete theoretical understanding of the observed
phenomena. A point in which the community working in this field has reached an 
overall agreement is the absence of any supersolid signal in a perfect
crystal, i.e., a crystal where the number of particles and the number of
sites matches exactly. The supersolidity is then attributed to alterations
of the perfect crystal due to the disorder introduced by defects, that
inevitably appear during its growing process from the liquid phase. The
major role has been assigned to dislocations and their mobility in relation
with the pining/depining of $^3$He impurities to them. But, even if
dislocations can be in the origin of experimental NCRI they can not be the
only explanation since the superfluidity of hcp $^4$He in bulk or in vycor
is nearly the same, whereas the dislocation density is reasonably expected
to be very different. Another possibility, that has been theoretically
explored from long time is the presence of a fraction of point defects,
i.e., vacancies.

Vacancies were originally proposed by Andreev and Lifshitz~\cite{Lifshitz1969} 
as a mechanism for creating a supersolid phase in solid helium. 
This scenario was rebuked by experimental and theoretical 
findings showing that vacancies are energetically too expensive to be created 
by \emph{thermal} activation.
However, Rossi \textit{et al.}~\cite{Reatto2008} showed using a variational estimate that
wavefunctions that are able to describe well the equation of state of
solid $^4$He support a finite vacancy concentration between 
$10^{-6}$ and $10^{-3}$. This possibility is yet
to be tested with first principles calculations, as the
number of helium atoms that corresponds to such low concentrations is
still too large to handle. Even such low vacancy concentrations should be 
experimentally significant, given the observed influence 
of sub-ppm concentrations of $^3$He~\cite{Chan2008}. Notice that as a 
quasiparticle in hcp $^4$He, 
a vacancy is in many ways similar to a $^3$He impurity.

Even if thermal activation of vacancies seems improbable at very low
temperature it is reasonable to consider that a significant number of vacancies
can be introduced into solid samples during their growth. 
Eventually, some of them can \textit{disappear} by migrating to
dislocations or grain boundaries but one cannot exclude a priori that a
tiny fraction of vacancies can dissolve into the bulk crystal.
It was, however, argued~\cite{Boninsegni2006} 
that even if initally vacancy-rich,
solid hcp $^4$He would phase separate into a vacancy-rich phase and a
perfect, insulating crystal and therefore any
growth-introduced vacancies will be effectively removed from the
experimental samples.
On the other hand, our detailed studies with several vacancies 
show no sign of vacancy clustering~\cite{Lutsyshyn2010b,Lutsyshyn2011}.

In this work, we report recent results on the properties of vacancies in
a fully quantum crystal like $^4$He obtained using different quantum
Monte Carlo methods. In Sec. 2, we describe the microscopic methods used in
the present analysis.  In Sec. 3, we
present PIMC results on the one-body density matrix of hcp solid $^4$He as
a function of temperature, showing the onset temperature where both Bose-Einstein
condensation and vacancy delocalization appear. Sec. 4 comprises results of
the shear modulus at zero temperature and as a function of the pressure.
Finally, an account of the main conclusions of the present work is included 
in Sec. 5.

\section{Quantum Monte Carlo methods in the study of solid $^{\bf 4}$He}

A quantitatively accurate study of solid $^4$He is demanding due to the
high density of the system and strong interparticle correlations. If a
microscopic approximation to the system is pursued, the most powerful tool
is quantum Monte Carlo with several methods to deal properly with zero or
finite temperature simulations. The final goal is to get relevant
microscopic information, on both energy and structure, starting directly from
the Hamiltonian of the system. The helium interatomic potential is
accurately known and there are several models that are able to describe
its equation of state very well. In the present work, we have used an Aziz
potential~\cite{aziz} that has proven to be very accurate in the
reproduction of the equation of state $P(\rho)$ in both liquid and solid
$^4$He.

Calculations at zero temperature have been  performed
with  diffusion Monte Carlo (DMC). DMC is a zero-temperature 
first-principles method which can access exactly the ground state of bosonic systems.
It is a form of Green's Function Monte Carlo
which samples the projection of the ground state from the initial configuration
with the operator $\exp{\left[-(\mathcal{H}-E_0)\tau\right]}$. Here, $\mathcal{H}$
is the system Hamiltonian, $E_0$ is a norm-preserving adjustable constant and $\tau$
is the variable which corresponds to imaginary time. The simulation is 
performed by advancing in $\tau$ via a combination of diffusion, drift and branching
steps on walkers (sets of $3N$ coordinates) representing the wavefunction of 
the system.

One of the advantages of DMC lies in the convenient incorporation of the
importance  sampling. The imaginary time evolution of the walkers is
``guided'' during the drift stage by a guiding wavefunction $\phi_G$, which
is usually a good guess for the  wavefunction of the system.  When
a guiding wavefunction is used, the expectation value of an operator
$\mathcal{A}$ computed with DMC results in a value equal to $\langle \phi_0
\left| \mathcal{A} \right| \phi_G \rangle$, where $\phi_0$ is the ground
state wavefunction of the system. This leads to a common misunderstanding 
regarding the
bias resulting from the choice of $\phi_G$. In fact, it is straightforward 
to show that for the Hamiltonian  $\mathcal{H}$ and any operator commuting with 
it, the expectation value  is computed exactly within statistical error.
That is, if $\left[\mathcal{A},\mathcal{H}\right]=0$, then
$\langle \phi_0 \left| \mathcal{A} \right| \phi_G \rangle = \langle \phi_0
\left| \mathcal{A} \right| \phi_0 \rangle$ for any $\phi_G$ such that
$\langle \phi_0 | \phi_G \rangle \ne 0$. In practice, one usually demands
that the DMC results are not sensitive to small   changes in parameters
describing $\phi_G$, if the parameters are sufficiently close to the optimal
values. The use of the importance sampling is the reason why DMC results
agree so precisely with   a wide range of energetic and structural 
experimental data, both for liquid and solid $^4$He.

The wavefunction that we use for the importance sampling of
solid helium is a symmetrization of the well-known 
Nosanow--Jastrow~\cite{Nosanow1964}
wavefunction. The symmetrized version has the form
\begin{equation}
\phi_\abbrev{snj}=\left(\prod_{i<j}^{N_{\text{p}}} f\left(\left|\bm{r}_i-\bm{r}_j\right|\right) \right)
\left( \prod_{k}^{N_{\text{s}  \makebox[0cm]{\phantom{$\scriptscriptstyle p$}}     }} 
\sum_i^{N_\text{p}} g\left(\left|\bm{r}_i-\bm{l}_k\right|\right)\right)\text{.}
\label{dmc-snj}
\end{equation}
The first term in $\phi_\abbrev{snj}$ is the McMillan product of pair
correlation  functions $f(r)=\exp\left[-1/2\,(b/r)^5\right]$.  The second
term describes the lattice structure. Here $\bm{r}$ and $\bm{l}$ describe
respectively positions of the $N_\text{p}$ atoms and $N_\text{s}$ lattice
sites. The localizing function $g(r)$ is set to a Gaussian,
$g(r)=\exp(-1/2\,\gamma r^2)$. The two parameters, $b$ and $\gamma$, are
obtained from optimization of  the unsymmetrized form of the function, as
described in Ref.~\cite{Lutsyshyn2010}. This allows for excellent
convergence, including insensitivity of results  with respect  to changes
in the parameters, as described above.

The symmetrized wavefunction of Eq.~(\ref{dmc-snj}) is an excellent and, at
the  same time, simple description of a quantum crystal. Already on the
variational level,  this wavefunction is capable of predicting the
solid-to-liquid transition for $^4$He (optimized $\gamma$ experiences a
discontinuous jump to $\gamma=0$, corresponding to the liquid phase), which
by itself is a very remarkable result.  The second lattice term in
$\phi_\abbrev{snj}$ is a product of sums over all particles. The product is
maximized when the smallest possible number  of the sums is vanishingly
small. This corresponds to largest possible number of lattice sites being
occupied.  The product therefore includes all possible  permutations of
atoms on the lattice, but the atoms are more likely to occupy maximum
number of sites in each permutation. Interstitials are allowed but have a
built-in probability penalty, as it should be. Exchanges between atoms are
allowed by construction.  Vacancies may be introduced by using
$N_\text{p}<N_\text{s}$, which only  changes the norm of
$\phi_\abbrev{snj}$. Structural defects may be created by perturbing the
set of the lattice sites $\{\bm{l}\}$. This wavefunction was first
introduced and tested with DMC for quantum solids  by Cazorla \textit{et
al.}~\cite{Boronat2009NJP} and has been  since then applied to a
range of problems including  elasticity~\cite{Cazorla2011}
and point defects~\cite{Lutsyshyn2011,Lutsyshyn2010}
in solid $^4$He, and even quantum crystals of  atoms excited to the Rydberg
states~\cite{Osychenko2011}.

We are also interested in the behavior of vacancies in solid $^4$He at
finite temperature. In this case, the appropriate theoretical tool is
the path integral Monte Carlo (PIMC) method. Starting from
the Hamiltonian $\mathcal{H}$ and the temperature $T = (k_B
\beta)^{-1}$ of the system, it is possible to rewrite the partition
function $Z$ making use of the convolution property that the thermal density
matrix $\rho({\bf R}',{\bf R};\beta)=\langle {\bf R}' \vert
e^{-\beta\hat{H}} \vert {\bf R} \rangle$ satisfies,
\begin{equation}\label{PartitionFunction}
Z = \rm{Tr}(e^{-\beta \mathcal{H} }) \simeq \int \prod_{i=1}^{M}dR_i \, 
\rho\left({\bf R}_i,{\bf R}_{i+1};\varepsilon\right) \ ,
\end{equation}
with $\varepsilon=\beta/M$, and the boundary condition $R_{M+1} = R_1$. 

PIMC describes the quantum $N$-body system
considering $M$ different configurations ${\bf R}_j$ of the same system,
whose sequence constitutes a path in imaginary time. 
This means that the $N$-body quantum system is mapped onto a
classical system of $N$ ring polymers, each one composed by $M$ beads. The
different beads can be thought as a way to describe the delocalization of
the quantum particle due to its zero-point motion.

For sufficiently large $M$, we recover the high-temperature limit for the
thermal density matrix, where it is legitimate to separate the kinetic
contribution from the potential one (primitive action). In this way, 
it is possible to reduce
the systematic error due to the analytical approximation for $\rho$ below
the statistical uncertainties and therefore to  recover ``exactly'' the
thermal equilibrium properties of the system. However, 
the primitive action is too simple for studying extreme quantum matter and a
better choice for the action is fundamental to reduce the complexity of
the calculation and ergodicity issues. Using the Chin action
\cite{chin,Sakkos09}, we are able to obtain an accurate  estimation of the
relevant physical quantities with reasonable numeric effort even in the low
temperature regime, where the simulation becomes harder  due to the large
zero-point motion of particles.

An additional problem we have to deal with when approaching the low
temperature limit with PIMC simulations arises from the indistinguishable
nature of $^4$He atoms. In the path integral formalism, the exchanges
between $L$ different particles are represented by long ring polymer
composed by $L \times M$ beads. If we study a bosonic system, the
indistinguishability of the particles does not affect the positivity of the
integrand function in Eq. \ref{PartitionFunction} and thus the symmetry of
$Z$ can be recovered via the direct sampling of permutations between the
ring polymers. A very efficient sampling scheme that we have used in the
present study is provided by the Worm Algorithm \cite{BoninsegniWorm}. 

The formation of long permutation cycles
is a frequent event at very low temperature and polymers which close themselves 
winding the
periodic boundary conditions of the simulation box can appear in the PIMC
configurations. The winding number, that is the net number of times the paths of
the polymers wind around the periodic cell, is an important quantity since
it is related to the superfluid fraction of the system \cite{Pollock87}. In
particular, the appearance of polymers presenting non-zero winding numbers
in PIMC configurations gives indication of superfluidity in the simulated system.

\section{Vacancies at finite temperature}

The presence of vacancies in solid $^4$He produces at zero temperature the
emergence of supersolidity, even when the vacancy concentration is very small.
The influence of vacancies in solid $^4$He is reasonably
expected to change with temperature, but an accurate estimation of
temperatures at which supersolidity appears was lacking until recently. 
Our aim has been to
gain a deeper understanding of this behavior by performing comprehensive PIMC
simulations. It is worth mentioning that the efficiency of the sampling in
PIMC, when the temperature approaches the zero limit, drops progressively
by the low acceptance ratio of the
sampling movements. In order to overcome, at least in part, these technical
issues it is crucial to work with an accurate action, that allows for
reducing the number of terms (beads), and a good permutation sampling as the
one provided by the worm algorithm.
  
We have studied the properties of incommensurate solid $^4$He  at finite
temperature  carrying out PIMC simulations of $N = 179$ $^4$He atoms,
interacting through an accurate Aziz pair potential~\cite{aziz}, in an
almost cubic simulation box matching the periodicity of an hcp lattice made
up of $N_s = 180$ sites at a density $\rho = 0.0294$ \AA$^{-3}$. As usual,
we apply periodic boundary conditions to the simulation box to emulate the  
infinite dimensions of the bulk system. 

Thanks to the accuracy of the Chin approximation for the action, it is possible 
to reach
convergence of the physical observables in the limit $\varepsilon \to 0$
with a rather large value of the imaginary time step $\varepsilon$, making
thus feasible the simulation of the quantum system with a small number of
beads, even at low temperature~\cite{Sakkos09}. More precisely, the convergence of the
one-body density matrix $\rho_1(r)$ is achieved with a time step $\varepsilon
= 0.033 \, \textrm{K}^{-1}$.

\begin{figure}
\begin{center}
\includegraphics[width=0.8\linewidth]{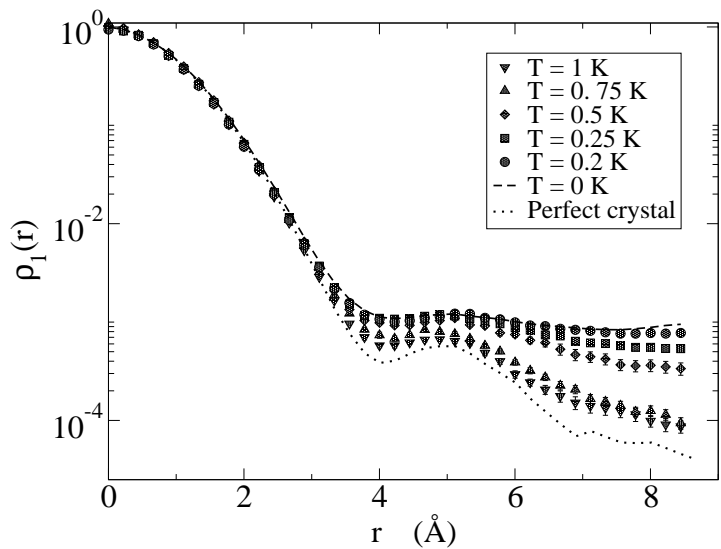}
\end{center}
\caption{The one-body density matrix $\rho_1(r)$
for an hcp crystal with vacancy concentration $X_v = 1/180$ at
density $\rho = 0.0294$ \AA$^{-3}$ and at different temperatures: $T =
1 \, {\rm K}$ (triangles down), $T = 0.75 \, {\rm K}$ (triangles up), 
$T = 0.5 \, {\rm K}$ (diamonds), $T = 0.25 \,
{\rm K}$ (squares) and $T = 0.2 \, {\rm K}$ (circles). The dotted and 
dashed lines represent $\rho_1(r)$
at zero temperature respectively for the commensurate ($X_v = 0$)
and incommensurate crystal ($X_v = 1/180$) at the same density,
taken from Ref. \cite{Rota11}.}
\label{obdm_vac}
\end{figure}

PIMC results for $\rho_1(r)$ at different temperatures and at the density quoted
above are shown in
Fig.~\ref{obdm_vac}. We have plotted, in the same figure, the zero-temperature 
estimation of $\rho_1$ for the same system and for a perfect
hcp crystal, obtained with the Path Integral Ground State 
method~\cite{Rota11}. We notice that the highest temperature at which the 
system
presents a non-zero condensate fraction $n_0$, indicated by a plateau in
the large $r$ behavior of $\rho_1(r)$, is $T_0 = 0.2 \, \textrm{K}$. Our
estimation for $n_0$ at this temperature is $n_0 = (8.4 \pm 0.8) \times
10^{-4}$. At temperatures $T > T_0$, the plateau disappears and the
one-body density matrix presents an exponential decay at large $r$, which
becomes more pronounced as the temperature increases, up to $T = 0.75 \,
\textrm{K}$. Above this temperature, the decay of $\rho_1(r)$ becomes
independent of $T$ and it is similar to the large $r$ behavior of
$\rho_1(r)$ in commensurate (perfect) crystals.

\begin{figure}
\includegraphics[width=\linewidth]{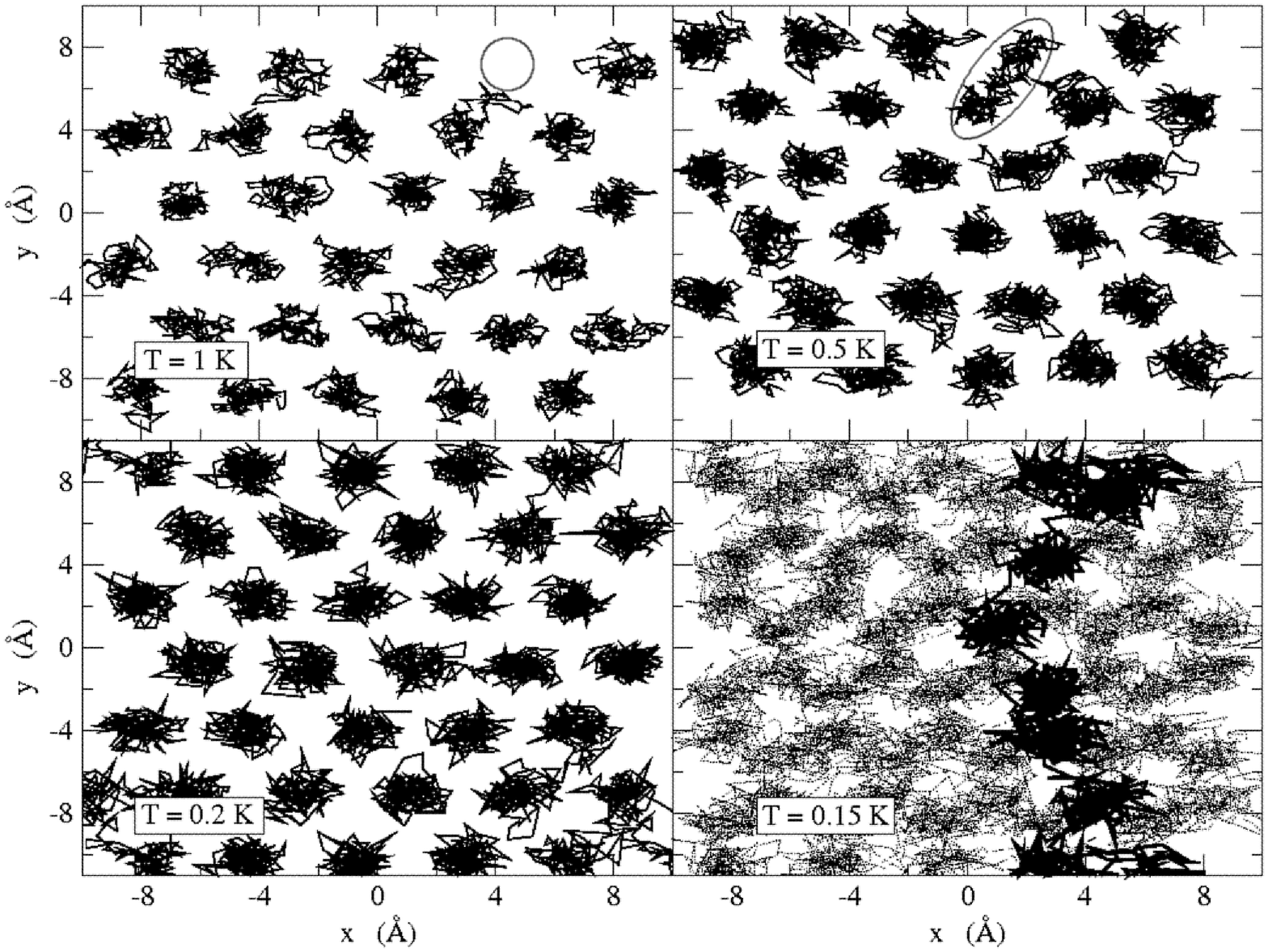}
\caption{Two-dimensional projection of basal planes
of the incommensurate hcp crystal at different temperatures,
represented according to the PIMC isomorphism of the classical
polymers. At $T = 1$ K (high-left panel) the vacancy is localized and
indicated by the circle. At $T = 0.5$ K (high-right panel), the
vacancy begins to delocalize: the ellipse indicate a quantum
particle delocalized over two different lattice sites. At $T =
0.2$ K (low-left panel), the vacancy is completely delocalized and
cannot be easily detected. Below $T = 0.2$ K (low-right panel), the 
delocalization of vacancy strongly enhance the exchange between the bosons 
and allows the appearance of paths presenting a non-zero winding number, 
like the one represented by the thick line.}\label{configurations}
\end{figure}

A relevant feature of PIMC is the possibility of giving a qualitative
microscopic description of atomic positions in $^4$He crystals by 
means of its ring-polymer representation; the spreading of beads of each
polymer gives an account of
the delocalization of the atoms due to their zero-point motion. In
Fig.~\ref{configurations}, we show snapshots of the typical configurations
of the incommensurate crystal at different temperature, plotting
two-dimensional projections of the PIMC polymers lying in a basal plane of
the hcp lattice.

At the highest temperature that we have studied, that is 
$T = 1 \, \textrm{K}$, the
polymers do not spread on distances larger than the interatomic distance,
indicating that the $^4$He atoms tend to stay localized around their
equilibrium positions. In this case, the vacancies are easily detectable 
inside the
lattice. This behavior explains the fact that, at this temperature, the
presence of vacancies does not affect noticeably the overall behavior of
$\rho_1$ which, for the incommensurate crystal, is similar to the one of
the perfect crystal.

At lower temperatures, the delocalization of $^4$He
atoms increases, allowing polymers to occupy different lattice points.  In
typical configurations at $T = 0.5 \, \textrm{K}$, we can detect
polymers which spread on distances higher than the interatomic distance.
This eventuality, which indicates that the vacancies begin to delocalize,
enhances the possibility for the different polymers to superpose and thus
to permute. However, at this temperature the polymers spreading on
different lattice points are still rare events and the formation of long
permutation cycles, which are necessary for BEC, is inhibited.

If we further decrease the temperature, the polymers spreading on different
lattice points become more frequent and it is impossible to associate a
well defined lattice point to every quantum particle, as in the case at
higher temperature. At $T = 0.2 \, \textrm{K}$, the typical configuration
of the incommensurate crystal looks like a commensurate system, indicating
that the vacancies at this temperature are completely delocalized and thus
undetectable inside the crystal. The frequent occurrence of a same lattice
point occupied by beads belonging to different polymers strongly enhances
the exchanges between the $^4$He atoms and allows also the creation of long
permutation cycles. In particular, for $T \le 0.2 \, \textrm{K}$, it is
possible to sample configurations presenting a non zero winding number, as
the one shown in the low left panel of Fig.~\ref{configurations}. In this
picture, we show two following basal planes of the incommensurate hcp
lattice and we highlight a path which winds around the boundary conditions
of the simulation box. The appearance of non-zero winding number paths in
the sampled configurations is a clear signal that the simulated
incommensurate crystal supports superfluidity at temperatures below $T_0 =
0.2 \, \textrm{K}$. On the other hand, this onset temperature is observed
to be an increasing function of the vacancy
concentration~\cite{rota_letter}.

\section{Vacancies and the elastic constants}

We have carried out a computational study of the elastic properties
of perfect (e.g. free of defects) and incommensurate solid $^{4}$He in the hcp 
structure based on the diffusion Monte Carlo approach. 
This zero-temperature study is intended to improve our understanding of the   
response of solid helium to external strain, and extends 
the work initiated by Pessoa \emph{et al.}~\cite{pessoa10}. 
In particular, we provide the zero-temperature dependence of $C_{44}$  
on pressure up to $\sim 110$~bar. 
This is a significantly higher pressure than previously considered 
both experimentally and theoretically.
Our results are compared to experimental 
data and other calculations when available and, as it will be shown 
later on, good agreement is generally found. The computational method that
we employ is fully quantum and virtually exact, that is, in principle 
only affected by statistical uncertainties.
In this sense, our study also represents an improvement with respect to 
previous zero-temperature first-principles work~\cite{pessoa10} 
based on variational Monte Carlo 
calculations (i.e., subject to bias stemming from the
choice of the trial wavefunction).        

For small strains, the zero-temperature energy of a crystal can be 
expressed as
\begin{equation}
E = E_{0} + \frac{1}{2} V_{0} \sum^{6}_{i,j = 1} C_{ij} s_{i} s_{j}~, 
\label{eq:energy}
\end{equation}  
where $V_{0}$ and $E_{0}$ are the volume and internal energy
of the undistorted solid, $\lbrace C_{ij} \rbrace$ the elastic constants 
and $\lbrace s_{i} \rbrace$ the strain components defined such 
that $s_{1}$, $s_{2}$ and $s_{3}$ are fractional increases in the $x$, 
$y$ and $z$ directed axes, and $s_{4}$, $s_{5}$ and $s_{6}$ angular 
increases of the $xy$, $xz$ and $yz$ angles.~\cite{wallace72,king70} 
The symmetry of the crystal under consideration defines
the number of elastic constants which are non-zero. 
In hcp crystals, this number reduces to five 
namely  $C_{11}$, $C_{12}$, $C_{33}$, $C_{13}$ and $C_{44}$,    
where $C_{44}$ is commonly known as the shear modulus.  
To calculate these elastic constants, is necessary
to compute the second derivative of the internal energy of the crystal 
with respect to the strain tensor $\sigma_{ij}$. For this, we express
the primitive hcp unit cell in terms of the translational vectors
\begin{eqnarray}
{\bf a_{1}} & = & a \left(+\frac{1}{2} {\bf i} + \frac{\sqrt{3}}{2} {\bf j}
\right) \\
{\bf a_{2}}  & = & a \left(-\frac{1}{2} {\bf i} + \frac{\sqrt{3}}{2} {\bf
j} \right) \\
{\bf a_{3}}  & = & c {\bf k} \ ,
\end{eqnarray}
where $a$ and $c$ are the lattice parameters 
in the basal plane and along the $z$ axis respectively, 
and ${\bf i}$, ${\bf j}$ and ${\bf k}$ correspond to the 
usual unitary Cartesian vectors, and two-atom basis set 
${\bf r_{1}} = \frac{1}{2}{\bf a_{1}} + \frac{1}{3}{\bf a_{2}} + \frac{2}{3}{\bf a_{3}}$ 
and ${\bf r_{2}} = ( 0 , 0 , 0)$.

In particular, the shear modulus $C_{44}$ quantifies the response 
of the hcp crystal to a heterogeneous strain in which the angle between 
by the $c$-axis and basal plane is tilted and the volume of the unit 
cell kept fixed. 
Such a shear deformation can be expressed as a transformation 
between sets of primitive translational vectors,  
$\lbrace {\bf a_{1}}, {\bf a_{2}} , {\bf a_{3}} \rbrace \to
\lbrace {\bf e_{1}}, {\bf e_{2}} , {\bf e_{3}} \rbrace$~,
where~\cite{king70} 
\begin{eqnarray}
{\bf e_{1}} &=& a \left(+\frac{1}{2} {\bf i} + \frac{\sqrt{3}}{2} {\bf j} + \frac{\epsilon}{2} {\bf k} \right) \nonumber \\
{\bf e_{2}} &=& a \left(-\frac{1}{2} {\bf i} + \frac{\sqrt{3}}{2} {\bf j} - \frac{\epsilon}{2} {\bf k} \right) \nonumber \\
{\bf e_{3}} &=& c{\bf k}~,
\label{eqn:c44vec}
\end{eqnarray}
$\epsilon$ being a dimensionless parameter. 
It follows that 
\begin{equation}  
C_{44} = \frac{1}{V_{0}}\left( \frac{\partial^{2} E}{\partial \epsilon^{2}} \right)_{V=V_{0}}~,
\label{eq:c44}
\end{equation}
where the equilibrium condition is fulfilled at $\epsilon = 0$.

The simulation box used in our perfect~(defected) pure 
shear calculations contains $200$~($199$) $^{4}$He atoms 
and was generated by replicating the hcp unit cell $5$ times 
along ${\bf e_{1}}$ and ${\bf e_{2}}$, and $4$ times along ${\bf e_{3}}$.
In proceeding so, hexagonal symmetry in our supercell
calculations is guaranteed by construction.
Periodic boundary conditions were imposed 
along the three directions defined by the 
edges of the non-orthorhombic simulation box. 
Prior to our shear modulus calculations, we determined 
the value of the equilibrium $c/a$ ratio at each volume.
We found that regardless of the pressure considered 
the optimal $c/a$ value was always $1.62(1)$.     
In order to express our $C_{44}~(V)$ results as a 
function of pressure we employed the equations of state     
$P(V)$ reported in Ref.~\cite{Lutsyshyn2010} (in 
both perfect and defected cases), which were deduced 
employing the DMC method and considering accurate 
finite-size corrections to the total energy.~\cite{cazorla08a} 
Additional details of our elastic constant 
calculations can be found in Ref.~\cite{Cazorla2011}.

\begin{figure}
\centerline{
\includegraphics[width=0.8\linewidth]{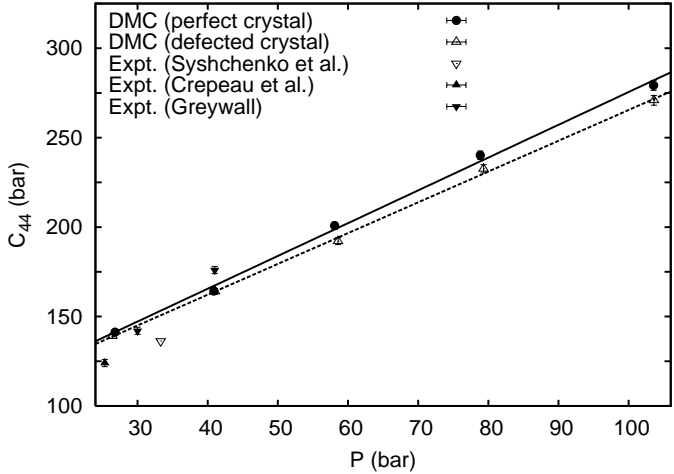}}%
\caption{Pressure-dependence of the calculated shear modulus of 
perfect and defected solid $^{4}$He at zero temperature. 
Experimental data are taken from Refs.~\cite{crepeau71} (Crepeau 
\emph{et al.}), \cite{greywall71} (Greywall), and 
\cite{syshchenko09} (Syshchenko \emph{et al.}) Linear
fits to our results are also shown.
}
\label{fig:c44-vac}
\end{figure}

\begin{table}[b]
\begin{center}
\begin{tabular}{l | c  c  c  c  c}
$\qquad \qquad $       & \qquad $C_{11}$ \qquad & \qquad $C_{12}$ \qquad  
& \qquad  $C_{13}$ \qquad  &  $\qquad C_{33}$ \qquad  &  
\qquad  $C_{44}$ \qquad  \\ 
\hline
Commensurate    &  $560(6)$  &  $210(3) $  &  $130(2) $  &  $639(7)$  &  $140(2) $ \\ 
Incommensurate  &  $574(6)$ &  $227(3)$  &  $152(2)  $  &  $649(2) $  & $138(2)$ \\ 
Expt.           &  $405(4)$  &  $213(4)$   &  $105(13)$  &   $554(22)$ & $124(2)$
\\
\end{tabular}
\end{center}
\caption{Elastic constants of the perfect (commensurate) and defected
(incommensurate) hcp solid $^4$He calculated with the DMC method at $P=26$ bar.
Experimental values obtained by Crepeau \textit{et al.}~\cite{crepeau71} at a
pressure $P=25.33$ bar. Numbers within parenthesis stand for errors. }
\end{table}

In Fig.~\ref{fig:c44-vac}, we plot the pressure dependence of the
shear modulus of perfect and defected solid $^{4}$He as obtained
in our $T = 0$ calculations. We found that these results can be 
accurately reproduced with a linear function of the 
form $C_{44}(P) = a_{44} + b_{44} P$.
In the perfect crystal, parameters $a_{44}$ and $b_{44}$ adopt
the values $92.2~(1.7)$~bar and $1.83~(0.04)$ respectively, whereas 
in the defected structure these are $93.3~(1.7)$~bar and $1.72~(0.04)$.    
It is observed that the elastic properties of perfect and defected
helium are very similar at pressures close to melting (that is $\sim 25$~bar). 
Also we note that both sets of $C_{44}$ data are compatible         
with experiments performed at low pressures and 
temperatures.~\cite{crepeau71,greywall71,syshchenko09}
A similar agreement is achieved with the other elastic constants that we report 
at a pressure close to melting in Table 1. At this particular pressure, all the
constants of the defected crystal except $C_{44}$ are larger than the ones 
of the perfect solid, with
differences of variable size depending on the particular $C_{ij}$.
As compression is increased, the shear modulus of the defected   
crystal becomes appreciably smaller than that of perfect $^{4}$He.
In relation to recent shear modulus experiments~\cite{syshchenko09} 
we must conclude that, based on this preliminary computational study,   
the increase of $C_{44}$ observed at low-$T$ can not be 
explained in terms of point defects.

\section{Summary and conclusions}

In the present work, we have reported recent results on the microscopic properties
of hcp $^4$He with a tiny fraction of point defects, i.e., vacancies. To make our
study as free of approximations as possible we have used state-of-the-art quantum
Monte Carlo methods, both at zero and finite temperature. Relying only on the
Hamiltonian, we have studied two particular aspects that we think relevant in the
present discussion about the possible supersolid $^4$He phase. First, we have
addressed the question of the behavior of vacancies when the temperature
decreases, approaching the temperatures where NCRI is observed. Our PIMC results
unambiguously show that there is an onset temperature for Bose-Einstein
condensation that depends on the vacancy concentration. For the lowest
concentrations that we can access with our approach this onset temperature is
close to the experimental one. At this same temperature, the vacancies are
observed to change their spatial behavior. Our simulations show that for $T$ larger
than the onset temperature, vacancies appear as classical entities in the sense that we can
always identify where they are, as it happens in a classical crystal. For
temperatures smaller than the onset one, the vacancy is completely delocalized and
the polymers representing the particles in the PIMC formalism spread over all the
sites as if the crystal were perfect.

In a second part, we have studied the elastic behavior of solid $^4$He in the
limit of zero temperature. Technically, this is achieved by calculating energies
of the deformed crystal by means of the DMC method. This approach is quite
standard in classical simulations but much less worked out in quantum Monte Carlo.
We have reported results for all the elastic constants at melting and the pressure
dependence of $C_{44}$. The presence of vacancies is also studied. Our preliminary
results show a globally small effect with the more relevant result being the
different slope of the elastic constants with the pressure: the slope for the
incommensurate crystal is systematically smaller than the one for the commensurate
phase.

The role of vacancies in a supersolid scenario for solid $^4$He has been
frequently excluded due to two arguments: the relatively high energy cost of its
formation  (above 10 K) and the possibility that vacancies form aggregates and eventually
evaporate. It is certainly difficult that vacancies appear at very low temperature
by thermal activation but we think that we cannot exclude the appearance of point
defects along the growing process. Concerning the possible aggregation of
vacancies due to a short-range attraction between them, there are different
theoretical predictions with contradictory observations. We are currently
exploring this issue as a function of temperature. 

\begin{acknowledgements}
The authors acknowledge partial financial support from the  
DGI (Spain) Grant No.~FIS2008-04403 and Generalitat de Catalunya 
Grant No.~2009SGR-1003.
\end{acknowledgements}

\pagebreak

\end{document}